\documentclass[12pt,preprint]{aastex}
\usepackage{graphicx}
\usepackage{epstopdf}

\newcommand{\solar}{\ifmmode_{\sun}\else$_{\sun}$\fi}

\newcommand{\HII}{H$\,${\sc ii}}
\newcommand{\HI}{H$\,${\sc i}}
\newcommand{\rknot}{$R_{knot}$}
\newcommand{\rd}{$R_D$}
\newcommand{\rbr}{$R_{Br}$}
\newcommand{\ratd}{$R_{knot}/R_D$}
\newcommand{\ratbr}{$R_{knot}/R_{Br}$}

\begin{document}

\title{Young star clusters in the outer disks of LITTLE THINGS dwarf irregular galaxies}

\author{
Deidre A. Hunter\altaffilmark{1},
Bruce G. Elmegreen\altaffilmark{2},
Elizabeth Gehret\altaffilmark{1}
}

\altaffiltext{1}{Lowell Observatory, 1400 West Mars Hill Road, Flagstaff, Arizona 86001 USA}
\altaffiltext{2}{IBM T. J. Watson Research Center, PO Box 218, Yorktown Heights, New York 10598 USA}

\begin{abstract}
We examine $FUV$ images of the LITTLE THINGS sample of nearby dwarf irregular (dIrr) and Blue
Compact Dwarf (BCD) galaxies to identify distinct young regions in their far outer disks. We
use these data, obtained with the {\it Galaxy Evolution Explorer} satellite, to determine the
furthest radius at which {\it in situ} star formation can currently be identified. The $FUV$
knots are found at distances from the center of the galaxies of 1 to 8 disk scale lengths and
have ages of $\le20$ Myrs and masses of 20 M\solar\ to $1\times10^5$ M\solar. The presence of
young clusters and OB associations in the outer disks of dwarf galaxies shows that dIrrs
do have star formation taking place there in spite of the extreme nature of the
environment. Most regions are found where the \HI\ surface density is $\sim1$ M\solar\
pc$^{-2}$, although both the \HI\ and dispersed old stars go out much further. This limiting
density suggests a cutoff in the ability to form distinct OB associations and perhaps even
stars. We compare the star formation rates in the $FUV$ regions to the average rates expected
at their radii and beyond from the observed gas, using the conventional correlation for
gas-rich regions.  The localized rates are typically 10\% of the expected average rates for the
outer disks. Either star formation in dIrrs at surface densities $<1\;M_\odot$ pc$^{-2}$ occurs
without forming distinct associations, or the Kennicutt-Schmidt relation over-predicts the rate
beyond this point. In the latter case, the stellar disks in the far-outer parts of dIrrs result
from scattering of stars from the inner disk.
\end{abstract}

\keywords{galaxies: irregular --- galaxies: star formation}

\section{Introduction} \label{sec-intro}

Star formation in dwarf irregular galaxies (dIrrs ) is difficult to understand because gas densities are low, but
this is particularly true in the outer parts where the gas densities are even lower than in the central regions.
There, the standard 2-dimensional \citet{toomre64} gravitational instability parameter for the gas,
suggested by \citet{kennicutt89} as an important parameter for star formation, predicts that the gas is highly
stable against collapse into star-forming clouds \citep{hunter11}. Yet, $FUV$ emission, a signpost of recent star
formation, is detected even into the outer disks.

In a deep-imaging study, we obtained $V$-band and {\it Galaxy Evolution Explorer} \citep[{\it GALEX};][]{GALEX}
$FUV$ images of 4 dIrr galaxies and one Blue Compact Dwarf (BCD), and $B$-band images for 3 of the
dIrrs \citep{hunter11}. Through surface photometry we traced stellar disks in these images to $\mu_V\sim 30$ mag
arcsec$^{-2}$. We found that the stellar surface brightness in both $V$ and $FUV$ continues exponentially as far
out as we can measure \citep[see also][]{bellazzini14}. The presence of $FUV$ emission into the outer disk
suggests a continuity of star formation with radius even though the outer gas is stable against
two-dimensional gravitational collapse in the conventional model \citep{bruce15}.

In spiral galaxies, numerical simulations show that spiral arms are capable of scattering stars into the outer
disks \citep{roskar08,minchev12,radburn12}, and this could be a significant source of the stellar populations there.
However, dIrrs do not have spiral arms. Are outer disks of dIrrs populated primarily through stars scattered from
the inside out or from {\it in situ} star formation? Are the outer parts of dIrrs from halos and not disks
\citep{martin14}? If star formation is taking place in the outer parts of dIrrs, then there should
be young $FUV$-bright star clusters, not just diffuse $FUV$ emission. We have previously identified clusters and
associations to a radius of a few disk scale lengths \citep{melena09}. Here we use $FUV$ images to search for
young star clusters and OB associations in the far outer regions of a larger sample of nearby dIrr
galaxies. Since scattering takes longer than the ages of young clusters and strong scattering may destroy
clusters \citep[but see][]{carraro06}, this search is a good test for {\it in situ} star formation in outer disks.

We also examine the environments for star formation in the far outer parts of these
galaxies (\S \ref{environ}). The pressure and average density are low, gravity is weak,
and the dynamical time is long. Most of the identified regions occur near the radius where the
average gas surface density drops below $\sim1\;M_\odot$ pc$^{-2}$, suggesting a cutoff or
slowdown at lower densities. The gas and distributed stellar population go further than this.
To test whether the outer disk might be built up from dispersed regions like those we observe,
we compare in \S {\ref{outer}} the star formation rates (SFRs) in the $FUV$ regions to the
average rates expected in the outer disks from the Kennicutt-Schmidt relation
\citep{kennicutt12}.

\section{Sample, Data, and Procedure} \label{sec-data}

The sample of galaxies is taken from LITTLE THINGS \citep[Local Irregulars That Trace
Luminosity Extremes, The \HI\ Nearby Galaxy Survey,][]{lt12}. This is a multi-wavelength survey
of nearby ($<10.3$ Mpc) dIrr galaxies and Blue Compact Dwarfs (BCDs), which builds on the
THINGS project, whose emphasis was on nearby spirals \citep{walter08}. The galaxies and a few
key parameters are listed in Table \ref{tab-data}. The galaxies Haro 29, Haro 36, Mrk 178, and
VIIZw 403, which appear at the end of the table, are classified as BCDs. NGC 3738, classified
as a dIrr, has characteristics, such as a central concentration of stars, star formation, and
gas, that are similar to BCDs, and so we will include it with the BCDs when that group is
singled out.

The LITTLE THINGS sample was chosen to be relatively isolated because we are investigating
internal processes of star formation. None of the galaxies are companions to
large spiral galaxies, members of dense clusters, or obviously engaged in an interaction with another galaxy.
Most are, however, members of groups and so other galaxies are around. In addition a few galaxies
and particularly the BCDs 
(NGC 1569, IC 10, Haro 29, Haro 36, Mrk 178, VIIZw 403)
are undergoing bursts of star formation that may have been triggered by some sort of external process, including mergers
\citep{ic10_13,ashley13,ashley14}. 
\citet{he04} gives the nearest neighbor to each galaxy and their separation.  
The non-starburst LITTLE THINGS galaxies are 110-1100 kpc from their nearest known neighbor, 
with a median separation of 480 kpc.
The starburst sub-sample is at a comparable range of distances -- 240-900 kpc with a median separation of 450 kpc.
Therefore, we expect that for most of the galaxies the extra-galactic environment is not a factor in
determining the extent of star-formation in the systems.

\begin{deluxetable}{lcccrrccc}
\tabletypesize{\scriptsize}
\tablewidth{0pt}
\tablecaption{Furthest FUV region \label{tab-data}}
\tablehead{
\multicolumn{4}{c}{} & \multicolumn{5}{c}{---------------------------Furthest $FUV$ region---------------------------} \\
\colhead{} & \colhead{D} & \colhead{$R_D$\tablenotemark{a}}
& \colhead{$R_{Br}$\tablenotemark{b}} & \colhead{RA} & \colhead{DEC}
& \colhead{$R_{knot}$\tablenotemark{c}} & \colhead{} & \colhead{} \\
\colhead{Galaxy} & \colhead{(Mpc)} & \colhead{(kpc)}
& \colhead{(kpc)} & \colhead{(h:m:s)} & \colhead{(d:am:as)}
& \colhead{(kpc)} & \colhead{$R_{knot}/R_D$}
& \colhead{$R_{knot}/R_{Br}$}
}
\startdata
%\cutinhead{Im Galaxies}
CvnIdwa   &  3.6 & 0.25$\pm$0.12 &  0.56$\pm$ 0.49 & 12:38:37.8 & +32:45:40 & 0.49$\pm$0.03 &  2.0$\pm$ 0.9 &    0.9$\pm$ 0.8 \\
DDO 43    &  7.8 & 0.87$\pm$0.10 &  1.46$\pm$ 0.53 &  7:28:16.7 & +40:47:01 & 1.93$\pm$0.08 &  2.2$\pm$ 0.3 &    1.3$\pm$ 0.5 \\
DDO 46    &  6.1 & 1.13$\pm$0.05 &  1.27$\pm$ 0.18 &  7:41:26.9 & +40:05:08 & 3.02$\pm$0.06 &  2.3$\pm$ 0.1 &    2.4$\pm$ 0.3 \\
DDO 47    &  5.2 & 1.34$\pm$0.05 &                 \nodata &  7:42:00.0 & +16:49:40 & 5.58$\pm$0.05 &  4.2$\pm$ 0.2 &   \nodata \\
DDO 50    &  3.4 & 1.48$\pm$0.06 &  2.65$\pm$ 0.27 &  8:18:23.8 & +70:43:18 & 4.86$\pm$0.03 &  3.3$\pm$ 0.1 &    1.8$\pm$  0.2 \\
DDO 52    & 10.3 & 1.26$\pm$0.04 &  2.80$\pm$ 1.35 &  8:28:26.3 & +41:50:25 & 3.39$\pm$0.10 &  2.7$\pm$ 0.1 &    1.2$\pm$  0.6 \\
DDO 53    &  3.6 & 0.47$\pm$0.01 &  0.62$\pm$ 0.09 &  8:34:11.8 & +66:10:12 & 1.19$\pm$0.03 &  2.5$\pm$ 0.1 &    1.9$\pm$  0.3 \\
DDO 63    &  3.9 & 0.68$\pm$0.01 &  1.31$\pm$ 0.10 &  9:40:56.0 & +71:12:15 & 2.89$\pm$0.04 &  4.2$\pm$ 0.1 &    2.2$\pm$  0.2 \\
DDO 69    &  0.8 & 0.19$\pm$0.01 &  0.27$\pm$ 0.05 &  9:59:12.4 & +30:44:37 & 0.76$\pm$0.01 &  4.0$\pm$ 0.2 &    2.8$\pm$  0.5 \\
DDO 70    &  1.3 & 0.44$\pm$0.01 &  0.13$\pm$ 0.07 & 10:00:05.4 &  +5:17:47 & 1.34$\pm$0.01 &  3.1$\pm$ 0.1 &   10.3$\pm$  5.5 \\
DDO 75    &  1.3 & 0.18$\pm$0.01 &  0.71$\pm$ 0.08 & 10:11:08.5 &  -4:39:07 & 1.38$\pm$0.01 &  7.7$\pm$ 0.4 &    1.9$\pm$  0.2 \\
DDO 87    &  7.7 & 1.21$\pm$0.02 &  0.99$\pm$ 0.11 & 10:49:25.1 & +65:30:41 & 4.23$\pm$0.07 &  3.5$\pm$ 0.1 &    4.3$\pm$  0.5 \\
DDO 101   &  6.4 & 0.97$\pm$0.06 &  1.16$\pm$ 0.11 & 11:55:36.4 & +31:31:18 & 1.23$\pm$0.06 &  1.3$\pm$ 0.1 &    1.1$\pm$  0.1 \\
DDO 126   &  4.9 & 0.84$\pm$0.13 &  0.60$\pm$ 0.05 & 12:27:14.1 & +37:08:01 & 3.37$\pm$0.05 &  4.0$\pm$ 0.6 &    5.6$\pm$  0.5 \\
DDO 133   &  3.5 & 1.22$\pm$0.04 &  2.25$\pm$ 0.24 & 12:32:48.9 & +31:31:42 & 2.20$\pm$0.03 &  1.8$\pm$ 0.1 &    1.0$\pm$  0.1 \\
DDO 154   &  3.7 & 0.48$\pm$0.02 &  0.62$\pm$ 0.09 & 12:53:58.4 & +27:07:20 & 2.65$\pm$0.04 &  5.5$\pm$ 0.2 &    4.3$\pm$  0.6 \\
DDO 167   &  4.2 & 0.22$\pm$0.01 &  0.56$\pm$ 0.11 & 13:13:20.6 & +46:19:32 & 0.70$\pm$0.04 &  3.2$\pm$ 0.2 &    1.2$\pm$  0.3 \\
DDO 168   &  4.3 & 0.83$\pm$0.01 &  0.72$\pm$ 0.07 & 13:14:31.4 & +45:54:12 & 2.25$\pm$0.04 &  2.7$\pm$ 0.1 &    3.1$\pm$  0.3 \\
DDO 187   &  2.2 & 0.37$\pm$0.06 &  0.28$\pm$ 0.05 & 14:15:54.9 & +23:03:34 & 0.42$\pm$0.02 &  1.1$\pm$ 0.2 &    1.5$\pm$  0.3 \\
DDO 210   &  0.9 & 0.16$\pm$0.01 &               \nodata   & 20:46:51.7 & -12:51:20 & 0.29$\pm$0.01 &  1.8$\pm$ 0.1 &    \nodata \\
DDO 216   &  1.1 & 0.52$\pm$0.01 &  1.77$\pm$ 0.45 & 23:28:39.7 & +14:44:47 & 0.59$\pm$0.01 &  1.1$\pm$ 0.0 &    0.3$\pm$  0.1 \\
F564-V3   &  8.7 & 0.63$\pm$0.09 &  0.73$\pm$ 0.40 &  9:02:55.6 & +20:04:31 & 1.24$\pm$0.08 &  2.0$\pm$ 0.3 &    1.7$\pm$  0.9 \\
IC 1613   &  0.7 & 0.53$\pm$0.02 &  0.71$\pm$ 0.12 &  1:05:08.9 &  +2:14:15 & 1.77$\pm$0.01    &  3.3$\pm$ 0.1 &     2.5$\pm$  0.4 \\
LGS 3     &  0.7 & 0.16$\pm$0.01 &  0.27$\pm$ 0.08 &  1:03:56.4 & +21:53:56 & 0.32$\pm$0.01    &  2.0$\pm$ 0.1 &    1.2$\pm$  0.4 \\
M81dwA    &  3.6 & 0.27$\pm$0.00 &  0.38$\pm$ 0.03 &  8:23:52.5 & +71:01:31 & 0.71$\pm$0.03 &  2.6$\pm$ 0.1&    1.9$\pm$  0.2 \\
NGC 1569  &  3.4 & 0.46$\pm$0.02 &  0.85$\pm$ 0.24 &  4:30:58.9 & +64:50:20 & 5.19$\pm$0.03 & 2.5$\pm$ 0.1 &    1.3$\pm$  0.4 \\
NGC 2366  &  3.4 & 1.91$\pm$0.25 &  2.57$\pm$ 0.80 &  7:29:36.2 & +69:16:22 & 6.79$\pm$0.03 &  3.6$\pm$ 0.5 &    2.6$\pm$  0.8 \\
NGC 3738  &  4.9 & 0.77$\pm$0.01 &  1.16$\pm$ 0.20 & 11:35:47.2 & +54:31:31 & 1.21$\pm$0.05 &  1.6$\pm$ 0.1 &    1.0$\pm$  0.2 \\
NGC 4163  &  2.9 & 0.32$\pm$0.00 &  0.71$\pm$ 0.48 & 12:12:09.2 & +36:10:27 & 0.47$\pm$0.03 &  1.5$\pm$ 0.1 &    0.7$\pm$  0.4 \\
NGC 4214  &  3.0 & 0.75$\pm$0.01 &  0.83$\pm$ 0.14 & 12:15:30.8 & +36:13:46 & 5.46$\pm$0.03 &  7.3$\pm$ 0.1 &    6.6$\pm$  1.1 \\
NGC 6822  &  0.5 & 3.16$\pm$0.10 &  0.46$\pm$0.16  & 19:43:34.1 & -14:33:42 & 4.61$\pm$0.00   &  1.5$\pm$ 0.1 &   10.0$\pm$3.5 \\
SagDIG    &  1.1 & 0.32$\pm$0.05 &  0.57$\pm$ 0.14 & 19:30:00.8 & -17:39:56 & 0.65$\pm$0.01    &  2.0$\pm$ 0.3 &    1.1$\pm$  0.3 \\
WLM       &  1.0 & 1.18$\pm$0.01 &  0.83$\pm$ 0.16 &  0:01:46.9 & -15:27:52 & 2.06$\pm$0.01       &  1.7$\pm$ 0.0 &    2.5$\pm$  0.5 \\
%\cutinhead{BCD Galaxies}
Haro 29   &  5.8 & 0.33$\pm$0.00 &  1.15$\pm$ 0.26 & 12:26:18.3 & +48:29:46 & 0.86$\pm$0.06 &  2.6$\pm$ 0.2 &    0.7$\pm$  0.2 \\
Haro 36   &  9.3 & 1.01$\pm$0.00 &  1.16$\pm$ 0.13 & 12:46:59.4 & +51:36:53 & 1.79$\pm$0.09 &  1.8$\pm$ 0.1 &    1.5$\pm$  0.2 \\
Mrk 178   &  3.9 & 0.19$\pm$0.00 &  0.38$\pm$ 0.00 & 11:33:33.7 & +49:13:31 & 1.45$\pm$0.04 &  7.6$\pm$ 0.2 &    3.8$\pm$  0.1 \\
VIIZw 403 &  4.4 & 0.53$\pm$0.02 &  1.02$\pm$ 0.29 & 11:28:01.9 & +78:59:50 & 0.33$\pm$0.04 &  0.6$\pm$ 0.1 &    0.3$\pm$  0.1 \\
\enddata
\tablenotetext{a}{$R_D$ is the disk scale length given by \citet{herrmann13}.}
\tablenotetext{b}{$R_{Br}$ is the radius at which the $V$-band surface brightness profile changes slope, as
given by \citet{herrmann13}.}
\tablenotetext{c}{Radius of the furthest $FUV$ knot from the center of the galaxy determined in the plane of the galaxy.
The galaxy disk geometry was determined from $V$-band images by \citet{he06}.
The uncertainty is assumed to be 2\arcsec.}
\end{deluxetable}

We used $FUV$ (1516 \AA) images obtained by {\it GALEX} \citep{melena09, hunter10, hunter11,zhang12}
to identify knots of emission in the outer disks of each galaxy. In order to better distinguish knots from the
wide-spread diffuse emission, we subtracted the stellar continuum from each $FUV$ image using the $V$-band image.
The $FUV$ image was geometrically transformed to match the pixel scale and dimensions of the $V$-band image.
The $V$-band image was scaled and Gaussian-smoothed so that, as well as possible, when it was subtracted
from the $FUV$ image, the diffuse emission disappeared.

In order to select $FUV$ knots that we were confident belonged to the galaxy, we used $FUV-NUV$
false-color pictures produced by the {\it GALEX} pipeline to pick out bright, blue knots that
stood out compared to the surrounding background/foreground populations. In particular we were
looking at the density of comparably bright and blue objects well beyond the galaxy. The shape
of the object is also a clue: is it resolved compared to a star or lumpy? As a training
exercise, we began with a few galaxies also studied by \citet{melena09}, and compared our
identifications of individual regions with theirs. The distance to the furthest knot from the
center of the galaxy in the plane of the galaxy, \rknot,  was determined by comparing the
location of the knot to ellipses superposed that were used for surface photometry. The major
axis and minor-to-major axis ratio of the ellipses were determined for the galaxy from the
$V$-band image. We then identified the annulus or ellipse boundary at which the $FUV$ knot
fell. The \rknot\ are given in Table \ref{tab-data}, along with the ratios of \rknot\ to the
disk scale length, \rd, and to the break radius, \rbr. The break radius is the radius at which
the exponential stellar surface brightness profile changes slope. Dwarf irregular galaxies,
despite having irregular and sporadic star formation, have well-behaved exponential stellar
disk profiles, even out to 6-10\rd\ \citep{saha10, hunter11, bellazzini14}. These
azimuthally-averaged profiles, like those in spirals, often show abrupt breaks, most often
dropping in brightness into the outer disk more steeply than in the interior disk (Type II
profiles) but occasionally dropping less steeply (Type III). Surface brightness profiles
without breaks (Type I) are actually rare. \rd\ and \rbr\ were taken from \citet{herrmann13}.
Two galaxies (DDO 47, DDO 210) do not have breaks in their exponential disk surface brightness
profiles.

\begin{figure}
\epsscale{0.8}
\plotone{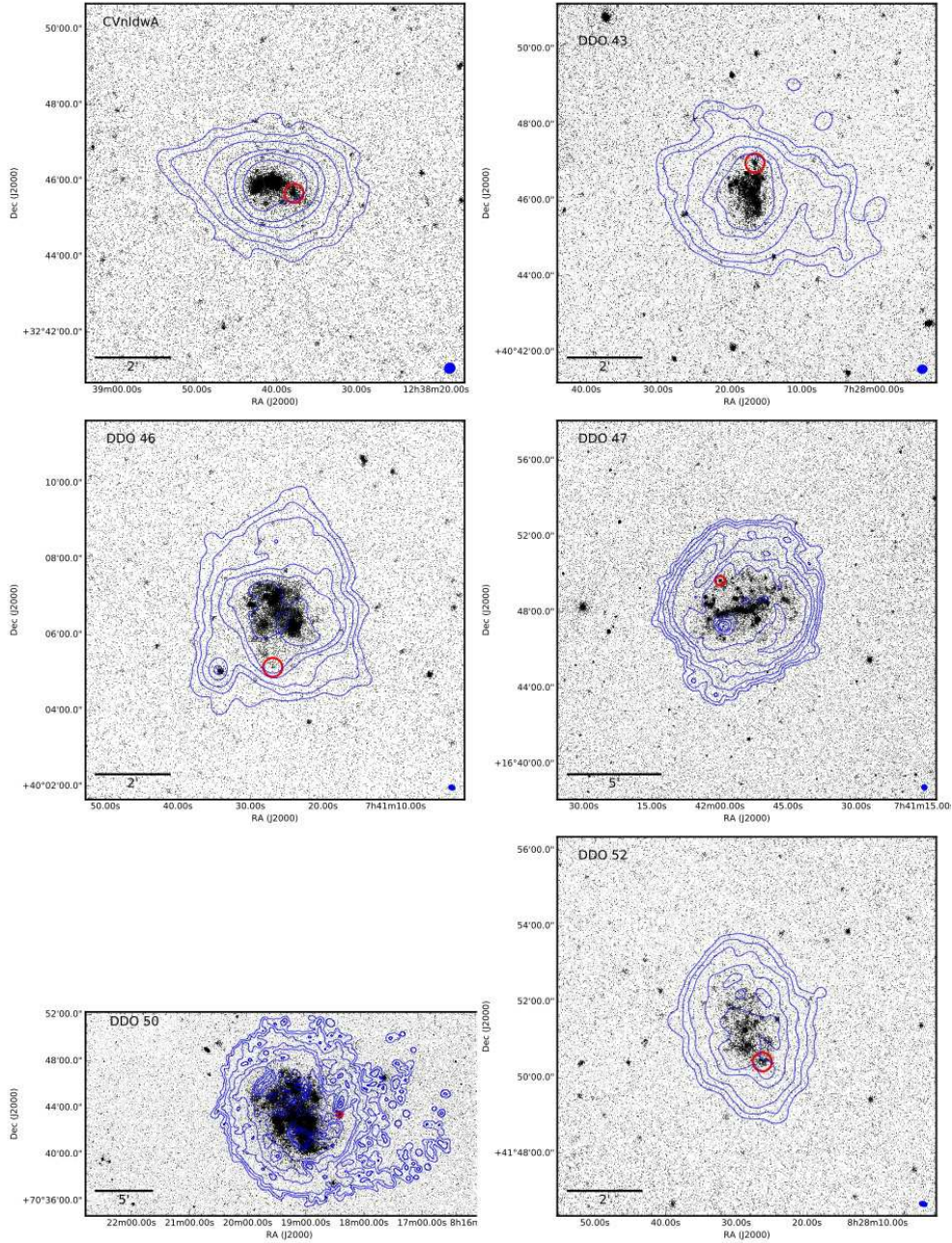}
\vskip -0.1truein
\caption{$FUV$ images of the galaxies in our sample. The red circle outlines the distinct $FUV$
region belonging to the galaxy that we considered to be at the furthest radius in the plane
of the galaxy. The circle has a radius of 15\arcsec\ unless the aperture used for
photometry of the region (given in Table \ref{tab-phot}) is larger; then the aperture
radius is used. The blue contours are of the integrated natural-weighted \HI\ map of the galaxy from
\citet{lt12} and outline column densities of 5, 30, 100, 300, 500, 1000, and
3000$\times10^{18}$ atoms cm$^{-2}$. The blue ellipse in the bottom right of each galaxy
panel outlines the FWHM of the \HI\ beam (major axis, minor axis, and position angle).
(The complete figure set (7 pages) is available in the online journal).                                                                        
\label{fig-bw}
}
\end{figure}

The $FUV$ regions are identified on the $FUV$ images shown in Figure \ref{fig-bw}, along
with the \HI\ shown as contours from the integrated \HI\ maps. 
The \HI\ extends much further than the outermost $FUV$ regions,
which are usually located just inside the 5th or 6th contour, counting inwards from the
outermost one. The \HI\ column density there is $5-10\times10^{20}$ atoms cm$^{-2}$, which
corresponds to $2.7-5.4\;M_\odot$ pc$^{-2}$ assuming a mean atomic weight of 1.36 times the
Hydrogen mass. The \HI\ maps are irregular, as expected for the dIrr morphology, and the $FUV$
regions are often associated with one of these irregularities, indicating the presence of a
local gas concentration. We show in \S \ref{relations} that the average \HI\ density at the
radius of the $FUV$ region is several times less than this local density. 

\section{Analysis} \label{sec-results}

\subsection{Characteristics of the $FUV$ regions}

We have estimated ages and masses of the knots by fitting aperture photometry from the $FUV$, $NUV$, $U$, $B$, and $V$ images \citep{he06}.
We determined the brightness of the underlying stellar disk for subtraction from
the region photometry for most of the regions using the mode of the counts in an annulus
5 pixels (1.35\arcsec\ to 5.67\arcsec) wide and 2 pixels (0.54\arcsec\ to 4.6\arcsec) in radius beyond the photometric aperture.
Since the regions are generally in the far outer parts of the disk, they are not crowded and aperture photometry is
adequate.
For some galaxies, such as those in which the star formation is centrally concentrated,
the $FUV$ regions are too crowded with other regions for an annulus around the region to give a
reliable measure of underlying galaxy. For these galaxies (DDO 53, DDO 101, M81dwA, NGC 3738, NGC 4163, Haro 29, Haro 36,
VIIZw 403), we determined the background from nearby areas at approximately the same disk surface brightness
as that where the $FUV$ region was found.
No $UBV$ photometry was available from our data for the $FUV$ region in NGC 6822.
The aperture radius, colors, and E($B-V$) used for reddening corrections are given in Table \ref{tab-phot}.

E($B-V$) includes foreground reddening from \citet{bh84} and 0.05 mag for internal reddening. The internal
reddening is half that determined from Balmer decrements in \HII\ regions in a sample of 39 dIrr galaxies, 
as described in \citet{hunter10}. The expectation is that these stars are not as embedded as
one would find in \HII\ regions.
Furthermore, since most of these regions are in the outer disks of the dwarfs, extinction
due to dust is expected to be minor.
We use the reddening law of \citet{cardelli89} to produce the extinction in each optical filter.
For the {\it GALEX} filters, we use $A_{FUV} = 8.24{\rm E(}B-V{\rm )}$ and $A_{NUV} = 7.39{\rm E(}B-V{\rm )}$ taken from
\citet{wyder07} and discussed in relation to the LITTLE THINGS dwarfs by \citet{hunter10}.

To model the ages and determine masses, we compared the integrated cluster photometry with cluster evolutionary models of
\citet{starburst99}. These models trace the evolution of a $10^6$ M\solar\ cluster over time, assuming
a \citet{salpeter55} stellar initial mass function from 0.1 to 100 M\solar.
We used the models for $Z = 0.004$ metallicity, since this is typical of the LITTLE THINGS dwarfs.
See \citet{hstcl02} and \citet{mccl03} for more details on our color fitting procedure.

The ages and masses of the clusters are given in Table \ref{tab-phot}, values are shown as histograms in
Figures \ref{fig-age} and \ref{fig-mass}, and mass is plotted against age in Figure \ref{fig-massvsage}.
The uncertainties in the ages and masses represent the range of ages
allowed by the colors. The ages are all less than 20 Myrs. Young ages are expected since the regions were
selected to be bright knots of $FUV$ emission. Masses range from 20 M\solar\ to $1.3 \times 10^5$ M\solar. There
is no correlation between age or mass and distance from the center of the galaxy.

Eight of the galaxies in our sample were also analyzed by \citet{melena09}. Working from the $NUV$ images,
they identified, not just the furthest regions, but all of the discrete regions in the galaxies. The ensembles of clusters
of these 8 galaxies extended to radii that were comparable to ours:
0.7-1.1 times the radii where we identified our furthest $FUV$ knots.
Their ensembles included older clusters, but the masses of our regions are within the range of
the masses found by \citet{melena09} out at their furthest radius.
The exceptions are DDO 70 and DDO 210 which have masses that are lower by a factor of 15 and 20, respectively, in our study.
Thus, we conclude that the regions we identified in our study are not unusual compared to the
collection of star clusters one finds in the outer disks of dIrr galaxies in general.

{
\begin{figure}[t!]
\epsscale{0.45}
\plotone{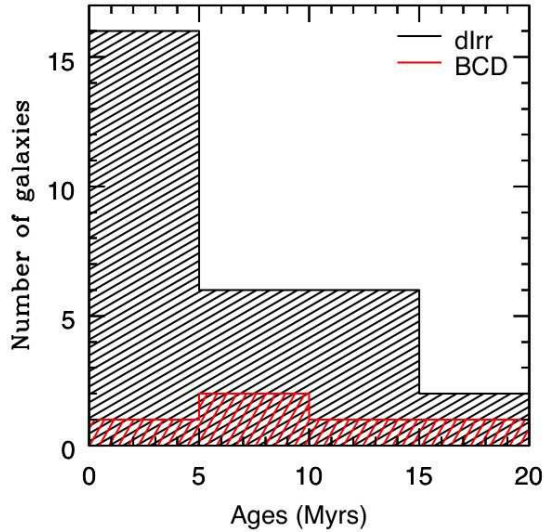}
%\vskip -0.25truein
\caption{
Histogram of the ages of the furthest $FUV$ knots found by fitting $FUV-NUV$, $U-B$, and $B-V$ colors
with Starburst99 cluster evolutionary models \citep{starburst99}.
\label{fig-age}}
\end{figure}

\begin{figure}[h!]
\epsscale{0.45}
\plotone{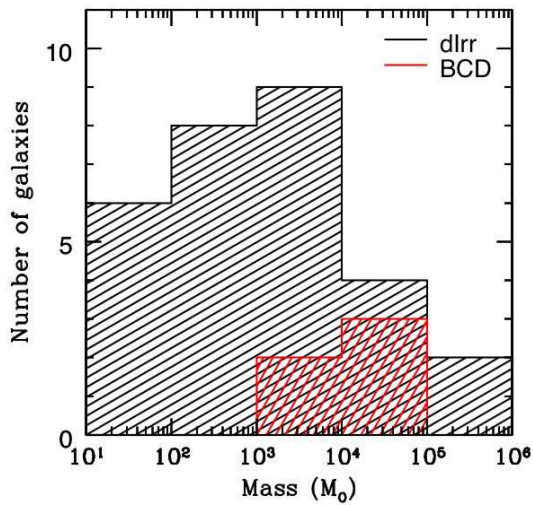}
%\vskip -0.25truein
\caption{
Histogram of the masses of the furthest $FUV$ knots, determined from the age
and $M_V$ using Starburst99 cluster evolutionary models \citep{starburst99}.
\label{fig-mass}}
\end{figure}
}

\clearpage

%%%New Figure 4
\begin{figure}[h!]
\epsscale{0.5}
\plotone{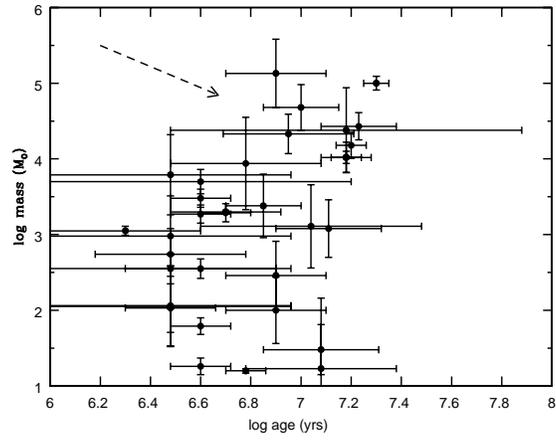}
\vskip 0.25truein
\caption{
Mass versus age of the $FUV$ knots.
The uncertainties in the ages and masses represent the range of ages allowed by the colors.
The dashed vector denotes the consequences of adding a reddening of E($B-V$)=0.1.
\label{fig-massvsage}}
\end{figure}

\clearpage

{
\begin{deluxetable}{lccccccccccc}
\tabletypesize{\scriptsize}
\rotate
\tablewidth{0pt}
\tablecaption{Photometry\tablenotemark{a}, ages and masses of $FUV$ regions \label{tab-phot}}
\tablehead{
\colhead{Galaxy} & \colhead{E(B-V)} & \colhead{r(arcs)\tablenotemark{b}}
& \colhead{$m_{FUV}$} & \colhead{$M_{FUV,0}$} & \colhead{$(FUV-NUV)_0$}
& \colhead{$M_{V,0}$} & \colhead{$(B-V)_0$} & \colhead{$(U-B)_0$}
& \colhead{Nfit\tablenotemark{c}} & \colhead{Age (Myrs)\tablenotemark{d}}
& \colhead{Mass (M\solar)}
}
\startdata
%\cutinhead{Im Galaxies}
CVnIdwA   & 0.055 &   5.7 & 20.786 &  -6.995$\pm$0.068 & -0.251$\pm$0.078 &  -6.982$\pm$0.071 & -0.001$\pm$0.089 & -0.627$\pm$0.142 & 3 &  3$\pm$ 2 &    360$\pm$  250 \\
DDO 43    & 0.105 &   5.7 & 21.009 &  -8.452$\pm$0.080 & -0.141$\pm$0.101 &  -8.771$\pm$0.122 & -0.355$\pm$0.138 & -0.802$\pm$0.074 & 3 &  5$\pm$ 1 &   2000$\pm$  490 \\
DDO 46    & 0.103 &   5.7 & 23.051 &  -5.876$\pm$0.229 & -1.178$\pm$0.611 &  -5.221$\pm$0.832 & -1.063$\pm$0.885 & -1.730$\pm$0.271 & 0 & \nodata & \nodata \\
DDO 47    & 0.073 &   7.9 & 20.711 &  -7.869$\pm$0.063 & -0.383$\pm$0.091 &  -8.045$\pm$0.122 & -0.123$\pm$0.147 & -0.663$\pm$0.100 & 3 &  3$\pm$ 2 &    960$\pm$  670 \\
DDO 50    & 0.073 &  14.7 & 18.804 &  -8.854$\pm$0.008 & -0.123$\pm$0.010 &  -8.475$\pm$0.061 & -0.203$\pm$0.068 & -1.078$\pm$0.049 & 3 &  4$\pm$ 1 &   1900$\pm$  440 \\
DDO 52    & 0.080 &   7.9 & 20.977 &  -9.087$\pm$0.082 & -0.422$\pm$0.118 &  -9.547$\pm$0.104 &  0.119$\pm$0.133 & -0.608$\pm$0.128 & 3 &  4$\pm$ 3 &   5000$\pm$ 1500 \\
DDO 53    & 0.075 &   5.0 & 19.402 &  -8.380$\pm$0.015 & -0.193$\pm$0.018 &  -8.787$\pm$0.003 &  0.022$\pm$0.004 & -0.914$\pm$0.004 & 3 &  5$\pm$ 2 &   2000$\pm$   90 \\
DDO 63    & 0.062 &  34.0 & 19.212 &  -8.743$\pm$0.013 &  0.065$\pm$0.016 &  -9.853$\pm$0.082 &  0.007$\pm$0.096 & -0.042$\pm$0.138 & 2 & 16$\pm$ 2 &  15000$\pm$ 4700 \\
DDO 69    & 0.050 &   9.1 & 21.245 &  -3.270$\pm$0.105 & -0.055$\pm$0.138 &  -3.362$\pm$0.116 &  0.099$\pm$0.131 & -0.766$\pm$0.130 & 3 & 12$\pm$ 5 &     30$\pm$   16 \\
DDO 70    & 0.063 &   5.7 & 21.486 &  -4.084$\pm$0.127 & -0.332$\pm$0.176 &  -3.466$\pm$0.299 & -0.213$\pm$0.351 & -1.008$\pm$0.185 & 3 &  4$\pm$ 1 &     18$\pm$    4 \\
DDO 75    & 0.068 &   4.7 & 20.101 &  -5.469$\pm$0.050 & -0.315$\pm$0.062 &  -5.664$\pm$0.049 &  0.320$\pm$0.067 & -1.113$\pm$0.063 & 2 &  3$\pm$ 1 &    110$\pm$   55 \\
DDO 87    & 0.050 &   6.8 & 21.938 &  -7.495$\pm$0.152 & -0.793$\pm$0.256 &  -7.439$\pm$0.184 &  0.514$\pm$0.262 & -0.744$\pm$0.312 & 1 &  3$\pm$ 3 &    550$\pm$   16 \\
DDO 101   & 0.058 &   6.8 & 20.447 &  -8.584$\pm$0.059 &  0.043$\pm$0.070 &  -9.519$\pm$0.006 &  0.308$\pm$0.010 & -0.806$\pm$0.013 & 1 & 15$\pm$ 3 &  10600$\pm$ 3900 \\
DDO 126   & 0.050 &  13.6 & 20.077 &  -8.374$\pm$0.054 & -0.169$\pm$0.070 &  -9.000$\pm$0.053 &  0.191$\pm$0.073 & -0.695$\pm$0.077 & 2 &  7$\pm$ 2 &   2400$\pm$ 1500 \\
DDO 133   & 0.053 &  11.3 & 18.827 &  -8.893$\pm$0.011 & -0.184$\pm$0.013 &  -8.992$\pm$0.020 &  0.159$\pm$0.025 & -1.002$\pm$0.023 & 2 &  4$\pm$ 1 &   3000$\pm$  710 \\
DDO 154   & 0.058 &  10.2 & 20.747 &  -7.094$\pm$0.087 & -0.858$\pm$0.153 &  -5.594$\pm$0.524 & -0.828$\pm$0.565 &  0.440$\pm$0.713 & 0 & \nodata & \nodata \\
DDO 167   & 0.050 &   4.5 & 21.541 &  -6.576$\pm$0.104 &  0.018$\pm$0.115 &  -7.320$\pm$0.096 &  0.394$\pm$0.116 & -0.626$\pm$0.099 & 2 & 13$\pm$ 5 &   1200$\pm$  710 \\
DDO 168   & 0.050 &   6.8 & 19.342 &  -8.825$\pm$0.036 &  0.176$\pm$0.038 & -11.396$\pm$0.008 &  0.507$\pm$0.015 &  0.725$\pm$0.037 & 1 & 20$\pm$ 2 & 100300$\pm$19000 \\
DDO 187   & 0.050 &   5.7 & 20.289 &  -6.423$\pm$0.050 & -0.232$\pm$0.054 &  -6.667$\pm$0.078 &  0.000$\pm$0.097 & -0.568$\pm$0.071 & 3 &  4$\pm$ 2 &    350$\pm$   90 \\
DDO 210   & 0.085 &   5.4 & 21.306 &  -3.465$\pm$0.098 & -0.616$\pm$0.148 &  -3.694$\pm$0.146 & -0.225$\pm$0.175 & -0.507$\pm$0.111 & 3 &  6$\pm$ 1 &     16$\pm$    1 \\
DDO 216   & 0.073 &   5.0 & 21.297 &  -3.910$\pm$0.095 & -0.122$\pm$0.108 &  -5.238$\pm$0.037 &  0.404$\pm$0.046 & -0.651$\pm$0.032 & 2 &  8$\pm$ 3 &    100$\pm$   60 \\
F564-V3   & 0.068 &   4.9 & 22.388 &  -7.310$\pm$0.172 & -0.450$\pm$0.262 &  -7.767$\pm$0.213 &  0.080$\pm$0.249 &  0.028$\pm$0.249 & 2 &  2$\pm$ 2 &   1100$\pm$  150 \\
IC 1613   & 0.055 &  23.2 & 18.606 &  -5.619$\pm$0.027 & -0.295$\pm$0.038 &  -5.718$\pm$0.111 & -0.006$\pm$0.121 & -0.565$\pm$0.096 & 3 &  3$\pm$ 2 &    110$\pm$   80 \\
LGS3      & 0.085 &   5.7 & 23.321 &  -0.905$\pm$0.395 &  0.020$\pm$0.518 &  -2.787$\pm$0.118 &  0.459$\pm$0.162 & -0.077$\pm$0.160 & 1 & 12$\pm$12 &     20$\pm$   15 \\
M81dwA    & 0.072 &   6.8 & 21.248 &  -6.534$\pm$0.026 & -0.049$\pm$0.033 & -11.477$\pm$0.002 &  0.927$\pm$0.004 &  0.675$\pm$0.010 & 1 & 10$\pm$ 3 &  48000$\pm$24000 \\
NGC 1569  & 0.558 &   4.2 & 22.336 &  -5.321$\pm$0.095 &  0.033$\pm$0.115 & -10.418$\pm$0.025 &  0.161$\pm$0.033 & -1.260$\pm$0.024 & 3 & 15$\pm$12 &  24000$\pm$18000 \\
NGC 2366  & 0.093 &  33.6 & 19.557 &  -8.101$\pm$0.051 & -0.149$\pm$0.070 & -10.489$\pm$0.020 &  0.590$\pm$0.025 & -0.135$\pm$0.023 & 1 &  6$\pm$ 3 &   8800$\pm$ 6700 \\
NGC 3738  & 0.050 &   4.5 & 16.175 & -12.276$\pm$0.008 & -0.099$\pm$0.009 & -13.063$\pm$0.001 &  0.161$\pm$0.002 & -0.706$\pm$0.002 & 2 &  8$\pm$ 3 & 130000$\pm$87000 \\
NGC 4163  & 0.050 &   4.5 & 18.736 &  -8.576$\pm$0.010 &  0.050$\pm$0.012 &  -9.520$\pm$0.002 &  0.324$\pm$0.004 & -0.579$\pm$0.009 & 2 & 15$\pm$ 2 &  11000$\pm$ 1800 \\
NGC 4214  & 0.050 &   6.3 & 21.373 &  -6.012$\pm$0.089 & -0.348$\pm$0.118 &  -5.738$\pm$0.146 & -0.230$\pm$0.157 & -0.770$\pm$0.071 & 3 &  3$\pm$ 2 &    110$\pm$   80 \\
NGC 6822  & 0.265 & 120.0 & 16.710 &  -6.785$\pm$0.016 & -0.239$\pm$0.021 & \nodata & \nodata & \nodata & 1 &  3$\pm$ 1 & \nodata \\
SagDIG    & 0.188 &   4.9 & 20.694 &  -4.513$\pm$0.083 & -0.117$\pm$0.104 &  -6.390$\pm$0.019 & -0.103$\pm$0.022 & -0.793$\pm$0.022 & 3 &  8$\pm$ 3 &    290$\pm$  180 \\
WLM       & 0.068 &   6.8 & 19.524 &  -5.476$\pm$0.041 & -0.275$\pm$0.051 &  -4.796$\pm$0.071 & -0.078$\pm$0.080 & -0.909$\pm$0.047 & 3 &  4$\pm$ 1 &     60$\pm$   14 \\
\tablebreak
%\cutinhead{BCD Galaxies}
Haro 29   & 0.050 &   4.9 & 19.518 &  -9.299$\pm$0.038 &  0.025$\pm$0.043 & -10.358$\pm$0.002 &  0.118$\pm$0.003 & -0.562$\pm$0.004 & 3 & 17$\pm$ 5 &  27000$\pm$ 9000 \\
Haro 36   & 0.050 &   4.9 & 19.777 & -10.065$\pm$0.042 & -0.079$\pm$0.050 & -10.833$\pm$0.003 &  0.179$\pm$0.005 & -0.857$\pm$0.008 & 2 &  9$\pm$ 4 &  21000$\pm$ 9400 \\
Mrk 178   & 0.050 &   6.7 & 20.271 &  -7.684$\pm$0.056 &  0.159$\pm$0.064 &  -7.478$\pm$0.049 & -0.260$\pm$0.051 & -0.992$\pm$0.029 & 2 & 11$\pm$ 7 &   1300$\pm$  900 \\
VIIZw 403 & 0.073 &   4.3 & 18.209 & -10.008$\pm$0.021 & -0.309$\pm$0.025 & -10.066$\pm$0.002 & -0.049$\pm$0.003 & -0.808$\pm$0.003 & 3 &  3$\pm$ 2 &   6100$\pm$ 4300 \\
\enddata
\tablenotetext{a}{$FUV$ magnitudes are AB magnitudes. The $UBV$ magnitudes are on the standard
\citet{landolt92} system.}
\tablenotetext{b}{Radius of photometry aperture in arcseconds.}
\tablenotetext{c}{Number of colors ($FUV-NUV$, $B-V$, $U-B$) used in the age determination. An Nfit of 1 means the $FUV-NUV$ color was used, a value of 2 means
either $U-B$ or $B-V$ was also used, and a value of 3 means a consistent age was determined for all three colors.}
\tablenotetext{d}{Uncertainties represent the range of ages allowed by the colors. Derivations of ages consistent with the observed colors were not possible for LGS3, WLM, and NGC 6822.}
\end{deluxetable}
}

\clearpage

\subsection{Locations of the $FUV$ regions}

\begin{figure}[t!]
\epsscale{0.5}
\plotone{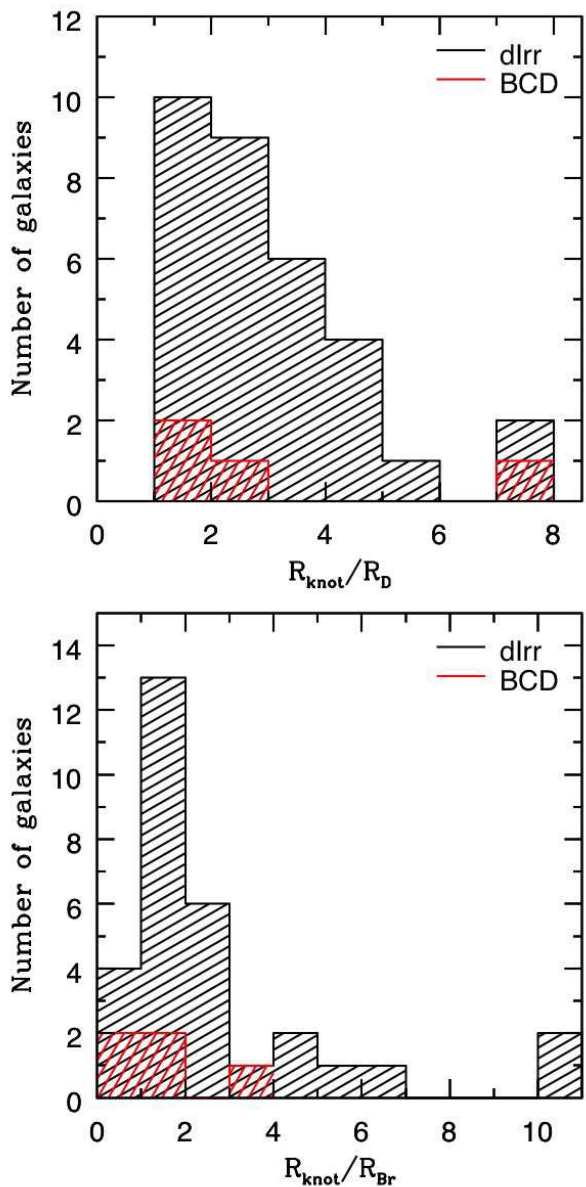}
%\vskip -0.25truein
\caption{
Histogram of the distance to the furthest knot of $FUV$ emission for each galaxy relative to the disk scale length \rd\ and the
break radius \rbr. The break radius is the radius at which the stellar exponential profile changes slope. Two dIrr galaxies in our sample
do not have breaks.
\label{fig-histrats}}
\end{figure}

Histograms of the ratios \ratd\ and \ratbr\ are shown in Figure \ref{fig-histrats}. There is structure due
to star formation present in almost all of the outer disks. All of the furthest $FUV$ knots in the dIrr galaxies
are beyond one disk scale length and 69\% are beyond 2\rd. A few (DDO 70 and NGC 4214) are out as far as 7-8\rd.
In terms of the break in the $V$-band surface brightness profiles, in 57\% of the dIrr galaxies the furthest
$FUV$ knots are at radii $\le$2\rbr, and so the regions are found roughly near the break. In two of the galaxies
(DDO 75 and NGC 6822) the regions are found around 10\rbr.

NGC 6822 is one of the extreme dIrr galaxies with a \ratbr\ ratio of 10. The $FUV$ region is located in an \HI\ cloud
to the northwest of the galaxy center at a distance of 4.6 kpc.
This region had been suggested to be a companion that is interacting with
NGC 6822 \citep{deblok00, Komiyama03},
but an {\it HST} study of the stellar populations of this region and others along the \HI\ extent
found that the star formation histories of all six of their positions
are similar over the past 500 Myr \citep{cannon12}.
\citet{cannon12} argue that the ``companion" has an old stellar population that
is like that of the extended halo of NGC 6822.
Therefore, we include it here as the furthest $FUV$ region in NGC 6822.

DDO 75 is also a dIrr with a \ratbr\ ratio of 10, but it also has a peculiar surface brightness profile.
The $V$-band surface brightness profile is flat to mildly increasing in
brightness from the center to 0.7 kpc, and then the surface brightness drops like a normal exponential profile to a radius
of 1.3 kpc \citep{herrmann13}. After that, as the surface brightness becomes fainter with radius, there are superposed up and down
wiggles in the surface brightness profile \citep{bellazzini14}.
Therefore, \rd\ may not be as well defined as it is in other galaxies.

The two dIrrs with the highest \ratd\ ratios are DDO 70 and NGC 4214. The surface brightness profile of DDO 70 is also flat
but only to a radius of 0.1 kpc and then is well-behaved after that \citep{bellazzini14}.
NGC 4214 has a steeper profile in the inner disk, to a radius of 0.8 kpc. Beyond that radius, the surface brightness drops as
a normal exponential.
In spite of these small peculiarities, it is not obvious what is special about DDO 70 and NGC 4214 that they would have $FUV$
knots at such large \rd;
there are other dwarfs that are more peculiar than these with smaller \ratd.

For the BCDs, 43\% of the furthest $FUV$ regions are beyond 2\rd\ and 80\% are found relatively near the break
radius, with only one further out at 3-4\rbr. Statistics are poor, of course, with only five BCDs. Nevertheless,
we expect BCDs to be more centrally concentrated than normal dIrrs, and their \rknot\ are roughly in agreement
with that. The BCDs in the LITTLE THINGS sample have also been shown to likely be the result of an interaction or
merger that has driven material towards the central regions of the galaxies \citep{ashley13, ashley14}.

\subsection{Relationships of $FUV$ regions to other galactic characteristics}
\label{relations}

\begin{figure}[t!]
\epsscale{0.5}
\plotone{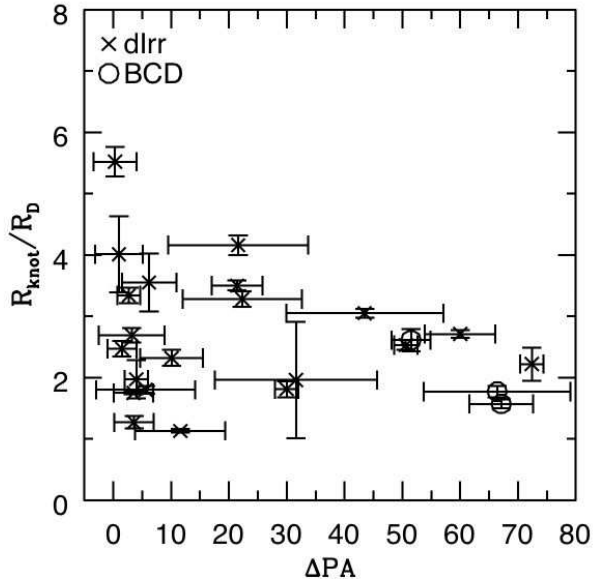}
\vskip -0.1truein
\caption{
\ratd\ plotted versus $\Delta$PA, the difference between the position angle of the major axis of the stellar disk, determined
from $V$-band images, and the position angle of the \HI\ kinematics, determined from fits to the bulk velocity field of the gas
\citep{oh15}.
We assume an uncertainty in the optically derived PA of 2\arcdeg\ and the uncertainty in the \HI\ kinematics is from \citet{oh15}.
\label{fig-delpa}}
\end{figure}

To explore a potential connection between large-scale structural peculiarities and the extended nature of
star formation in dwarfs, we look at the relationship between \ratd\ and differences between optical and \HI\ kinematic
position angles. The position angle of the optical major axis was determined from smoothed outer isophotes of the $V$-band images
\citep{he06}, and the \HI\ kinematic major axis was determined from iterative fits to the bulk velocity field of the
gas \citep{oh15}. Twenty-six of the LITTLE THINGS galaxies were in the study by \citet{oh15} and all but two are included
in this study. We would expect these two position angles - one describing the stellar disk and one describing the gas disk -- to be
co-located, but over half of the LITTLE THINGS galaxies have $\Delta$PA that are larger than 20\arcdeg .
The large differences could represent some peculiarity of the galaxy structure.
In Figure \ref{fig-delpa} we plot \ratd\ as a function of the difference in the position angles.
We find that galaxies with large $\Delta$PA are not distinguished in terms of \ratd\ compared to those with small $\Delta$PA,
although there is a decline in the upper envelope such that \ratd\ declines as $\Delta$PA increases.

\begin{figure}[t!]
\epsscale{0.5}
\plotone{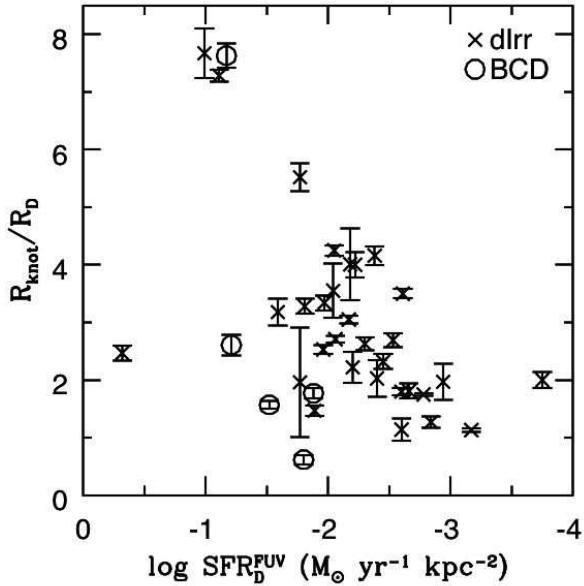}
\vskip -0.1truein
\caption{
\ratd\ plotted versus the log of the star formation rate determined from the integrated $FUV$ luminosity \citep{hunter10, zhang12}.
The SFR is normalized to the area of the galaxy within one disk scale length: $\pi R_D^2$.
The uncertainties in the SFR$_D^{FUV}$ are smaller than the point size and are not plotted.
\label{fig-sfr}}
\end{figure}

\begin{figure}[h!]
\epsscale{0.5}
\plotone{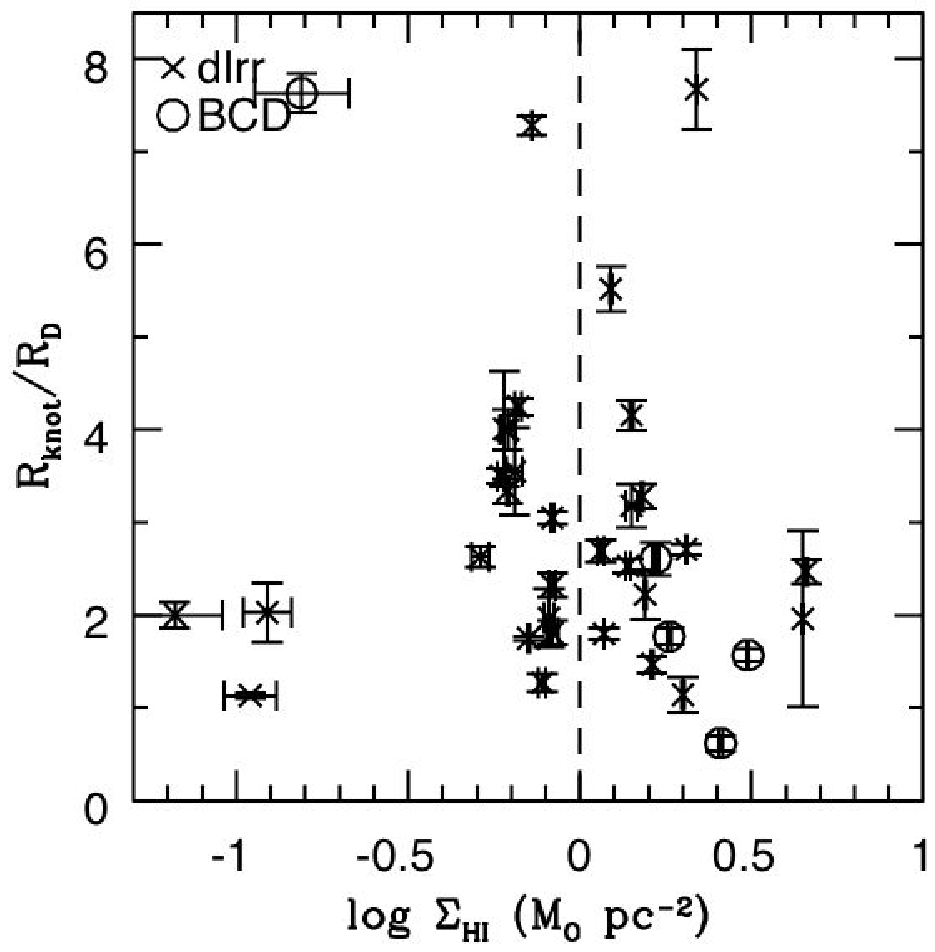}
\vskip -0.1truein
\caption{
\ratd\ plotted versus the log of the azimuthally-averaged
\HI\ surface density \citep{lt12} at the radius of \rknot.
The uncertainties in $\Sigma_{HI}$ are determined from the rms in the integrated \HI\ map and the number of beams
within the annulus over which an average was taken. The rms in the integrated \HI\ map was determined
from the uncertainty in individual channels and an assumed average number of channels contributing to each pixel
\citep[see][]{deblok08}.
\label{fig-hi}}
\end{figure}

In Figure \ref{fig-sfr} we plot \ratd\ as a function of the galactic integrated SFR determined from
the $FUV$ luminosity \citep{hunter10, zhang12}:
$\log {\rm SFR}_D^{FUV}$ (M\solar yr$^{-1}$ kpc$^{-2}$).
The SFR is normalized to the area contained within one disk scale length.
The thought here is that if a galaxy has a higher SFR, it might also have more extended star formation.
In Figure \ref{fig-sfr} we see that, with one exception, the galaxies with the highest SFRs do also have the furthest $FUV$
knots in terms of disk scale lengths. In addition, the galaxies with lowest SFRs have among the lower \ratd\ ratios.
However, the galaxies in between these extremes display a range of SFRs and a range of \ratd\ without a clear trend.

To explore the relationship between the extent of the $FUV$ knots and the average 
gas densities at that radius, we plot \ratd\
against the log of the azimuthally-averaged
\HI\ surface density $\Sigma_{HI}$ \citep{lt12} at the radius of \rknot\
in Figure \ref{fig-hi}. There is no trend of \ratd\ with $\Sigma_{HI}$, but most points are centered around
a value of 1 M\solar pc$^{-2}$.

\section{Environment for star formation near the knots}
\label{environ}

\begin{figure}
\epsscale{0.7}
\plotone{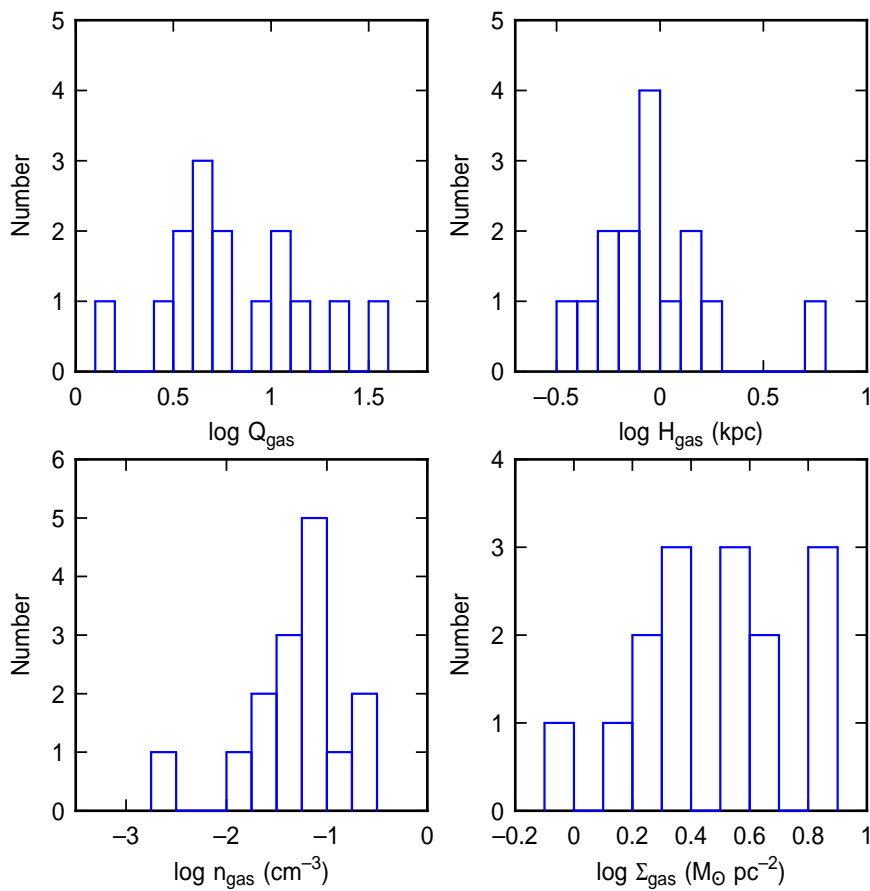}
\caption{Distribution of environmental parameters for 15 $FUV$ knots in the present
study that have measurements in the same galaxy and radial range considered
in \citet{bruce15}.  The parameters are all for the gas: Toomre $Q$, scale height
$H_{\rm gas}$, midplane density $n_{\rm gas}$, and surface density including
He and heavy elements.
\label{fig-hisgas}}
\end{figure}

\begin{figure}
\epsscale{0.7}
\plotone{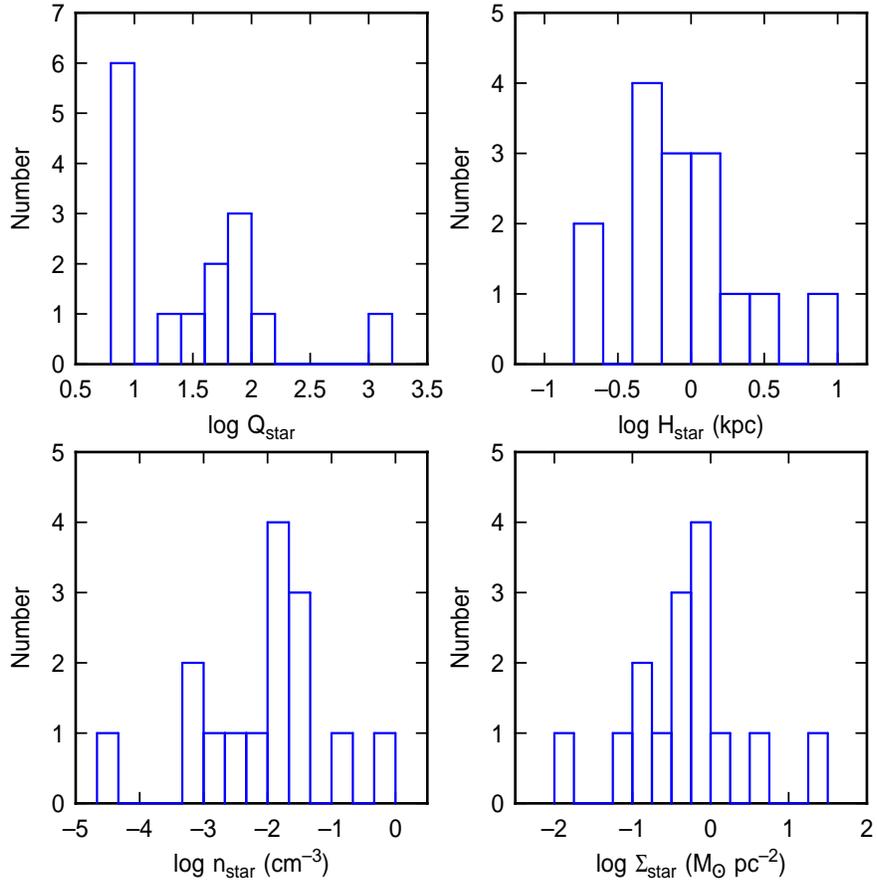}
\caption{Distribution of environmental parameters as in Figure \ref{fig-hisgas}
but for stars at the radius of the $FUV$ knots.
\label{fig-hisstar}}
\end{figure}

Fifteen galaxies in the present study have outer star formation knots within the radial
range of the survey of dIrr in \cite{bruce15}, for which we determined gas column
densities, scale heights, midplane densities, and Toomre $Q$ values as a function of radius
for gas and stars. The scale heights and densities came from solutions to the vertical
equilibrium of a three component fluid consisting of gas, stars, and dark matter, using
observed velocity dispersions and column densities. The $Q$ values also included the
observed rotation curves to give the epicyclic frequency.  Figures \ref{fig-hisgas} and
\ref{fig-hisstar} show histograms of these quantities for gas and stars, respectively, at
the radii of the $FUV$ knots. Comparing the two figures, the average densities and
column densities for the stars are much lower than for the gas at these radii,
emphasizing that the $FUV$ knots are in gas-dominated regions. $Q$ is high for both gas and
stars, but higher for stars because of the lower stellar surface densities. The scale
heights at these radii are about the same for the two components.

The average environment at the radius of an $FUV$ region is important for understanding
how the gas first got concentrated into a cloud to begin the process of star formation. \HI\
excesses near the identified regions are sometimes directly evident as deformations in the \HI\
contours in Figure \ref{fig-bw}. Almost all of the $FUV$ knots occur in larger-scale $FUV$
structures too. These large-scale structures typically have a size comparable to the disk scale
length. Their formation would presumably involve processes that operate on the local average
dynamical time, which depends on the average midplane density.  In these far-outer regions,
the average midplane density is typically 10 or more times lower than the gas density in the
solar neighborhood (which is about 1 cm$^{-3}$) as a result of both the low \HI\ surface density
and the large gas scale height from weak gravity (Fig. \ref{fig-hisgas}). Consequently, the
dynamical time over the region that the star-forming gas came from was long: $t_{\rm ff}=140$
Myr for midplane gas density $n_{\rm gas}$ at the typical value of $0.1$ cm$^{-3}$. The free
fall time is taken to be $t_{\rm ff}=(3\pi/32G\rho)^{1/2}$ for $\rho=1.36n_{\rm gas}m_{\rm H}$
and $m_{\rm H}=1.67\times10^{-24}$ g. This time is 20 times the average age of the knots, which
is 7 Myr from Table 2.  Unless there is a relatively dense layer of undetected molecular gas
near the midplane or some triggering event like an extragalactic cloud impact, the formation of
the observed knots by large-scale dynamical processes requires a relatively long time. This
suggests there was a threshold density or column density for the cluster to form so the cloud
could grow without early disruption from the first few stars.

A threshold surface density for star formation in the outer parts of dIrr galaxies is
most likely not the same as that in the solar neighborhood \citep[e.g.,][]{lada10} because the
pressures and metallicities in dIrrs are much lower. The threshold column density for strong
self-gravity in a cloud scales as the square root of the boundary pressure, with collapse
proceeding exponentially slower as $P/(G\Sigma_{\rm gas}^2)$ increases \citep{padoan12}. With
$P$ lower by a factor of 10, the threshold column density for strong gravity in our case should
be lower by a factor of about 3. That would make it $\Sigma_{\rm thresh}\sim 40\;M_\odot$
pc$^{-2}$ inside the star-forming region of a cloud, if we take the solar neighborhood
threshold as $116\;M_\odot$ pc$^{-2}$ from \cite{lada10}. Another threshold involves opacity
and the formation of strong coolants, like CO. The opacity threshold, measured in terms of a
mass column density threshold, scales inversely with metallicity, and the metallicities for
these dIrrs are lower than solar by a factor of about 8. Thus the mass column density for an
opacity threshold should be 8 times larger in these dIrrs than in the solar neighborhood. If
$\sim1.5$ mag is the local threshold for CO formation \citep{glover12}, then the corresponding
surface density of gas would be $230 \;M_\odot$ pc$^{-2}$ for the clouds that formed our $FUV$
knots. This is what we observe for CO emission in the dIrr WLM, where the metallicity has this
value \citep{rubio15}. If both strong self-gravity and high extinction are required for star
formation, then the larger of these two thresholds should apply. However, if strong
self-gravity alone is required and the opacity threshold automatically follows during the
ensuing collapse, then only the lower threshold is required. This cloud threshold,
$\sim40\;M_\odot$ pc$^{-2}$, is a factor of $\sim13$ larger than the average surface density of
\HI\ at the radius of the $FUV$ knots, which is $\sim3\;M_\odot$ pc$^{-2}$ from Figures
\ref{fig-hi} and \ref{fig-hisgas}. A factor of 13 increase in column density corresponds to a
factor of $13^{1/2}\sim3.6$ contraction in two dimensions. This is about the same factor
that clouds in the solar neighborhood would have to contract if the cloud threshold for star
formation always scales with the square root of the ambient pressure. In this sense, star
formation in outer dIrr disks may not be qualitatively different from other star formation.

This picture of star formation at the local dynamical rate changes if a particular event
was triggered by something rapid, such as an extragalactic cloud impact.  Some BCDs and
extremely low-metallicity dwarf galaxies have single dominant star-forming regions with lower
metallicities than in the rest of the galaxy \citep{jorge13,jorge14,jorge15}.  These regions
could be the sites of extragalactic cloud impacts with low metallicities from nearly pristine
gas. One example, Kiso 5639 \citep{elmegreen16}, has a star formation rate surface density in a
$\sim800$ pc region that is $\sim5\times$ larger than the value expected from the
Kennicutt-Schmidt relation at the likely gas surface density. The total young mass there is
$\sim10^6\;M_\odot$ for a $\sim10^8\;M_\odot$ galaxy.  These regions are all much larger and
brighter than the $FUV$ regions studied here, which look more typical for disk galaxies,
although they formed at extremely low gas densities.

\section{Does star formation occur in the far-outer regions?}
\label{outer}

One of the motivations for the present study was to determine if star formation could account
for the outer $FUV$ disks of dIrr galaxies, as opposed to stellar scattering from the inner disk. 
Here we compare
the star formation rate inside each $FUV$ knot, obtained from the ratio of the knot mass to the
age, to the total rate inside an annulus of width $R_{\rm D}$ at the radius of the knot, as
determined from the average Kennicutt-Schmidt (KS) relation for dIrrs at the local gas surface
density \citep{bruce15}.  If the total rate in the small knot is comparable to the total
rate expected from the gas in the annulus, which is based on large-scale averages from many
other observations, then one might conclude that star-forming regions like these can generally
populate outer disks. If the annular rate is much larger than the local rate, then either we
are missing $FUV$ regions that are too small, faint, or dispersed to detect
\citep[e.g.,][]{pellerin08}, or the average KS relation over-estimates the star formation rate
at and below a gas surface density of $\sim1\;M_\odot$ pc$^{-2}$, where our $FUV$ regions begin
to disappear.

Such an over-estimate would imply that star formation virtually stops below $\sim1\;M_\odot$
pc$^{-2}$, according to our understanding of conventional processes
\citep[e.g.,][]{kennicutt12}. In that case, stellar scattering from the inner disk would seem
to be important in filling the outer disk with stars. The outer disks have relatively smooth
exponential radial profiles in $V$-band down to $\sim0.1\;M_\odot$ pc$^{-1}$ or lower
\citep{herrmann13}. This is below the average stellar surface density at the radius of the
$FUV$ regions, as shown in Figure \ref{fig-hisstar}. 

\begin{figure}[t]
\epsscale{0.9}
\plotone{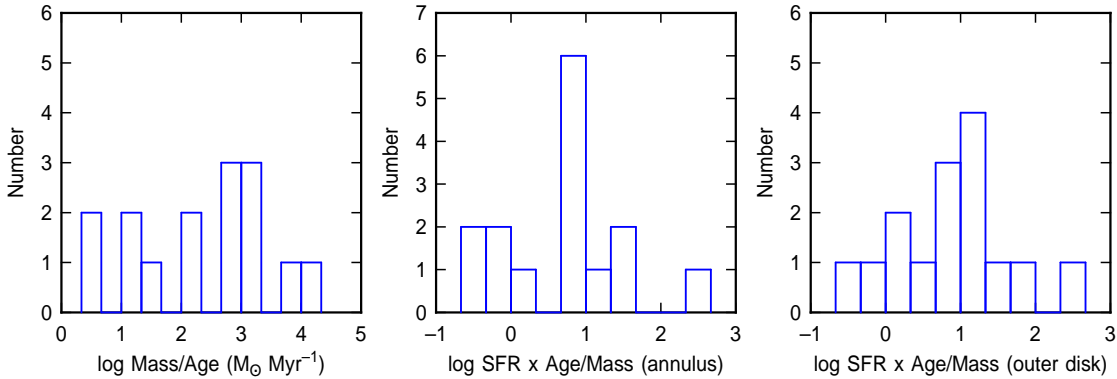}
\caption{(left) The distribution of the ratio of mass divided by age for the $FUV$ knots,
as an approximation to the average star formation rate. (middle) The ratio of the total
theoretical star formation rate in an annulus of width $R_{\rm D}$ centered on the knot
to the average rate in the knot. (right) The ratio of the total
theoretical star formation rate at all radii equal to and larger than the radius of the
knot to the average rate in the knot. The theoretical rate comes from the Kennicutt-Schmidt relation in
equation \ref{eq:ks} assuming an exponential gas disk with a scale length that is twice $R_{\rm D}$.
\label{fig-hissfr}}
\end{figure}

The left-hand panel of Figure \ref{fig-hissfr} shows a histogram of star formation rates in the
$FUV$ knots. There is a wide range of rates, primarily because the regions vary a lot in mass
and age. The middle panel shows a histogram of the ratio of the theoretical star formation rate
in an annulus of width $R_{\rm D}$ around a knot to the rate in the knot. The right-hand panel
shows the ratio of the theoretical rate integrated from the knot radius to infinity compared to
the knot rate. For the theory, we use the expression
\begin{equation}
\Sigma_{\rm SFR}=1.7\times10^{-5}\Sigma_{\rm gas}^2\; M_\odot\;{\rm Myr}^{-1}\;{\rm pc}^{-2}
\label{eq:ks}
\end{equation}
for \HI\ surface density $\Sigma_{\rm gas}$ in $M_\odot$ pc$^{-2}$. This implies that in the annulus,
\begin{equation}
SFR = 2\pi R_{\rm knot}R_{\rm D}\Sigma_{\rm SFR} \;M_\odot\;{\rm Myr}^{-1}
\end{equation}
and for the integral
\begin{equation}
SFR = 2\pi R_{\rm D}\left(R_{\rm D}+R_{\rm knot}\right)\Sigma_{\rm SFR} \;M_\odot\;{\rm Myr}^{-1}.
\label{eq:int}
\end{equation}
The theoretical rate follows from the expression $\Sigma_{\rm SFR}=\epsilon_{\rm ff}\Sigma_{\rm
gas}/t_{\rm ff}$ for a pure gas disk after substituting $\rho=\Sigma_{\rm gas}/(2H)$ and
$H=\sigma^2/(\pi G\Sigma_{\rm gas})$ inside the equation for free fall time given in the
previous section, i.e.,
\begin{equation}
1/t_{\rm ff}=4G\Sigma_{\rm gas}/[3^{1/2}\sigma].
\label{eq:tff}
\end{equation}
The theoretical rate compares well with observations of 20 dIrr galaxies if $\epsilon_{\rm
ff}=0.01$ and the velocity dispersion is the observed value of $\sigma\sim6$ km s$^{-1}$
\citep{bruce15,elmegreen15}. It also agrees with observations in the outer parts of spiral
galaxies, which are gas-dominated like the dIrrs \citep{elmegreen15}. This is the steep
part of the KS relation, beyond the main disks of spirals, where atoms dominate molecules, gas
dominates stars, and the total consumption time approaches 100 Gyrs \citep{bigiel08,bigiel10}.
Note that equation (\ref{eq:tff}) gives $t_{\rm ff}=590$ Myr for $\Sigma_{\rm gas}=1\;M_{\odot}$
pc$^{-2}$ and $\sigma=6$ km s$^{-1}$, so the gas consumption time would be $t_{\rm
ff}/\epsilon=59$ Gyr with $\epsilon=0.01$. For equation (\ref{eq:int}), we assume that the gas
disk has an exponential radial profile with a scale length twice that of the $V$-band disk, as
found for the 20 dIrrs and for the outer parts of 23 blue spirals by \citet{wang14}.

\begin{figure}
\epsscale{0.5}
\plotone{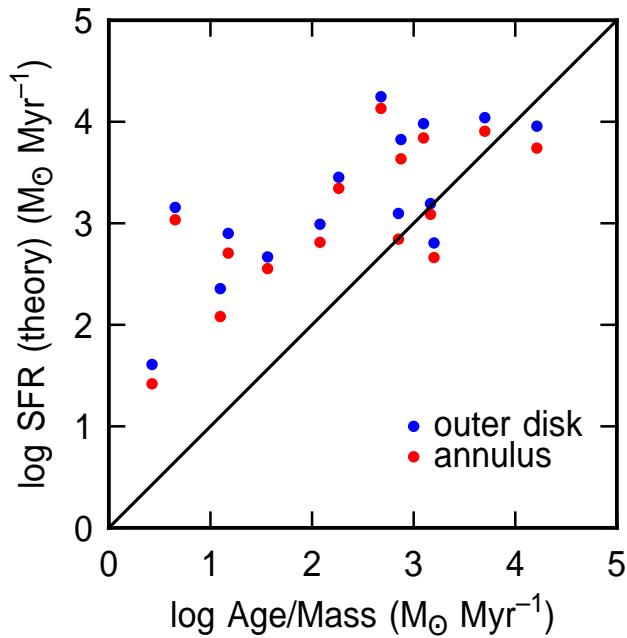}
\caption{The theoretical star formation rates are plotted versus the average rate in the
corresponding $FUV$ knot. The blue points for the outer disk integral are slightly higher
than the red points for an annulus around the knot. The line is the unity relation.
The rates are equal for the largest
knots, suggesting that star-forming regions like these could build up the outer disk.
The smallest knots cannot account for the outer disks in their galaxies, but we could be
missing similar knots because of faintness, or these particular knots could be uncharacteristically
small for their galaxies.
\label{fig-hissfr1}}
\end{figure}

The middle and right-hand panels in Figure \ref{fig-hissfr} suggest that the annuli around the
knots and the outer disks would have $\sim8\times$ larger star formation rates than the
knots have now, given the KS relation for average rates extrapolated to the local average gas
surface densities around the knots. Figure \ref{fig-hissfr1} illustrates this point by
plotting the two theoretical star formation rates for each knot location versus the knot's star
formation rate. Red dots are for the annulus and blue dots are for the outer disk. For the
largest knots the two rates agree well, but for the smallest knots, the star formation rates
fall short of the outer-disk expectations by a factor of $\sim10$.  As mentioned above, this
shortfall could be because smaller, fainter and more dispersed regions cannot be seen in our
data, or because the SFR is actually below the extrapolated KS relation at $\sim1\;M_\odot$
pc$^{-2}$.

\section{Summary} \label{sec-summary}

We have identified on {\it GALEX} images the furthest out distinct $FUV$ knot in the LITTLE
THINGS sample of dIrr and BCD galaxies that we are confident belong to the galaxy. These knots
are found at distances from the center of the galaxies of 1 to 8 disk scale lengths and have
ages of $\le20$ Myrs and masses of 20 M\solar\ to $1\times10^5$ M\solar. The presence of young
clusters in the outer disks of dwarf galaxies shows that dIrrs do have {\it in situ} star
formation taking place in their outer disks. Most regions are found around an \HI\ surface
density of 1 M\solar\ pc$^{-2}$, making this look like a threshold for the formation of
prominent OB associations.

The environments of these remote star-forming regions are extreme compared to the solar
neighborhood and inner Milky Way (\S \ref{environ}), or even compared to the inner parts of
dIrr galaxies. The average pressures and midplane densities are lower by a factor of $\sim10$,
the disks are relatively thick compared to the radial extents
\citep{hodge66,vdbergh88,bruce15}, and the metallicities are $\sim8$ times lower. Consequently,
self-gravity is weak in the interstellar medium, cloud contraction times are long, $\sim100$
Myr or more, and molecules are sparse. Moreover, the threshold column density for strong
gravity inside clouds should be $\sim3$ times lower than it is locally, placing this collapse
threshold firmly in the optically thin regime where atomic gas should still dominate.
Presumably molecules form after the collapse begins \citep{glover12,krumholz12}.

The star formation rates in the most remote $FUV$ knots fall short by a factor of $\sim8$
of the average rates obtained from the Kennicutt-Schmidt relation for these types of regions.
Only for the most massive $FUV$ knots are the rates comparable. Either we are missing
other discrete star-forming regions owing to faintness or rapid dispersal, or the star
formation rate drops below the KS relation for outer disks starting at around $\sim1\;M_\odot$
pc$^{-2}$.  In the latter case, stellar scattering from the inner disk would have to fill the outer disks with
stars, which go out much further than the last $FUV$ knot seen here.

\acknowledgments We are grateful to the Lowell Observatory Research Fund for funding,
including a summer internship for EG. 
We are also grateful to John and Meg Menke for a donation to Lowell Observatory 
that covered part of the page charges and to IBM for covering the remainder.
DAH also appreciates the considerable help on
installing and running python to produce Figure 1 from Joe Llama (University of St.
Andrews), Michael Mommert (NAU), and David Trilling (NAU).
We appreciate helpful comments by the referee.

Facilities: \facility{GALEX}

%\bibliography{my_bib}

\begin{thebibliography}{}
\bibitem[Ashley et al.(2013)]{ashley13} Ashley, T., Simpson, C. E., \& Elmegreen, B. G. 2013, \aj, 146, 42
\bibitem[Ashley et al.(2014)]{ashley14} Ashley, T., Elmegreen, B.G., Johnson, M., Nidever, D.L., Simpson, C.E.,
    Pokhrel, N.R.  2014, \aj, 148, 130
\bibitem[Bellazzini et al.(2014)]{bellazzini14} Bellazzini, M., Beccari, G., Fraternali, F., et al. 2014, A\&A, 566, 44
\bibitem[Bigiel et al.(2008)]{bigiel08} Bigiel, F., Leroy, A., Walter, F.,
    Brinks, E., de Blok, W. J. G., Madore, B., \& Thornley, M. D. 2008, AJ,
    136, 2846
\bibitem[Bigiel et al.(2010)]{bigiel10} Bigiel, F., Leroy, A., Walter, F., Blitz, L., Brinks,
    E., de Blok, W.J.G., Madore, B. 2010, AJ, 140, 1194
\bibitem[Billett et al.(2002)]{hstcl02} Billett, O. H., Hunter, D. A., \& Elmegreen, B. G. 2002, \aj, 123, 1454
\bibitem[Burstein \& Heiles(1984)]{bh84} Burstein, D., \& Heiles, C. 1984, \apjs, 54, 33
\bibitem[Cannon et al.(2012)]{cannon12} Cannon, J. M., O'Leary, E. M., Weisz, D. R., et al. 2012, \apj, 747, 122
\bibitem[Cardelli et al.(1989)]{cardelli89} Cardelli, J. A., Clayton, G. C., \& Mathis, J. S. 1989, \apj, 345, 245
\bibitem[Carraro et al.(2006)]{carraro06} Carraro, G., Villanova, S., Demarque, P., McSwain, M. V., Piotto, G.,
\& Bedin, L.\ R.\  2006, \apj, 643, 1151
\bibitem[de Blok \& Walter(2000)]{deblok00} de Blok, W. J. G., \& Walter, F. 2000, \apj, 537, L95
%\bibitem[de Blok \& Walter(2006)]{deblok06} de Blok, W. J. G., \& Walter, F. 2006, AJ, 131, 343
\bibitem[de Blok et al.(2008)]{deblok08} de Blok, W. J. G., Walter, F., Brinks, E., Trachternach, C., Oh, S.-H., \& Kennicutt, R. C., Jr. 2008, \aj, 136, 2648
\bibitem[Elmegreen(2015)]{elmegreen15} Elmegreen, B.G. 2015, \apj, 814, L30
\bibitem[Elmegreen \& Struck(2013)]{elmegreen13c} Elmegreen, B.G., \& Struck, C., 2013, \apj, 775, L35
\bibitem[Elmegreen \& Hunter(2015)]{bruce15} Elmegreen, B. G., \& Hunter, D. A. 2015, \apj, 805, 145
\bibitem[Elmegreen et al.(2016)]{elmegreen16} Elmegreen, D.M., Elmegreen,
    B.G., S\'anchez Almeida, J., Mu\~noz-Tu\~n\'on, C., Mendez-Abreu, J., Gallagher, J.S.
    Rafelski, M., Filho, M., \& Ceverino, D. 2016, ApJ, submitted
\bibitem[Glover \& Clark(2012)]{glover12} Glover, S.C.O. \& Clark, P.C. 2012, MNRAS, 421, 9
\bibitem[Herrmann et al.(2013)]{herrmann13} Herrmann, K.\ A., Hunter, D.\ A., \& Elmegreen, B.\ G. 2013, \aj, 146, 104
\bibitem[Hodge \& Hitchcock(1966)]{hodge66} Hodge, P. W., \& Hitchcock, J. L. 1966, PASP, 78, 7
\bibitem[Hunter \& Elmegreen(2004)]{he04} Hunter, D.\ A., \& Elmegreen, B.\ G. 2004, \aj, 128, 2170
\bibitem[Hunter \& Elmegreen(2006)]{he06} Hunter, D.\ A., \& Elmegreen, B.\ G. 2006, \apjs,162, 49
\bibitem[Hunter et al.(2003)]{mccl03} Hunter, D. A., Elmegreen, B. G., Dupuy, T. J., \& Mortonson, M. 2003, \aj, 126, 1836
\bibitem[Hunter et al.(2010)]{hunter10} Hunter, D. A., Elmegreen, B. G., \& Ludka, B. C. 2010, \aj, 139, 447
\bibitem[Hunter et al.(2011)]{hunter11} Hunter, D. A., Elmegreen, B. G., Oh, S.-H., et al. 2011, \aj 142, 121
\bibitem[Hunter et al.(2012)]{lt12} Hunter, D. A., Ficut-Vicas, D., Ashley, T., et al. 2012, \aj, 144, 134
\bibitem[Kennicutt(1989)]{kennicutt89} Kennicutt, R. C., Jr. 1989, \apj, 344, 685
\bibitem[Kennicutt \& Evans(2012)]{kennicutt12} Kennicutt R. C., \& Evans N. J., 2012, ARA\&A, 50, 531
\bibitem[Komiyama et al.(2003)]{Komiyama03} Komiyama, Y., Okamura, S., Yagi, M., et al. 2003, \apj, 590, L17
\bibitem[Krumholz(2012)]{krumholz12} Krumholz, M.R. 2012, \apj, 759, 9
\bibitem[Lada et al.(2010)]{lada10} Lada, C.J., Lombardi, M., \& Alves, J.F. 2010, \apj, 724, 687
\bibitem[Landolt(1992)]{landolt92} Landolt, A. U. 1992, \aj, 104, 340
\bibitem[Leitherer et al.(1999)]{starburst99} Leitherer, C., Schaerer, D., Goldader, J. D., et al. 1999, \apjs, 123, 3
\bibitem[Martin et al.(2005)]{GALEX} Martin, D.\,C., Fansom, J., Schiminovich, D., et al. 2005, \apj, 619, L1
\bibitem[Martin-Navarro et al.(2014)]{martin14} Mart\'in-Navarro, I., Trujillo, I., Knapen, J.H.,
    Bakos, J., Fliri, J. 52  2014, MNRAS, 441, 2809
\bibitem[Melena et al.(2009)]{melena09} Melena, N. W., Elmegreen, B. G., Hunter, D. A., \& Zernow, L. 2009, \aj,
    138, 1203
\bibitem[Minchev et al.(2012)]{minchev12} Minchev, I., Famaey, B., Quillen, A. C., Di Matteo, P., Combes, F.,
    Vlajic, M., Erwin, P., Bland-Hawthorn, J. 2012, A\&A, 548, A126
\bibitem[Nidever et al.(2013)]{ic10_13} Nidever, D., Ashley, T., Slater, C. T., et al.
%Ott, J., Johnson, M., Bell, E. F., Stanimirovi\'c S., Putman, M., Majewski, S. R., Simpson, C., Burton, W. B. 
2013, ApJL, 779, L15
\bibitem[Oh et al.(2015)]{oh15} Oh, S.-H., Hunter, D. A., Brinks, E., et al. 2015, \aj, 149, 180
\bibitem[Padoan et al.(2012)]{padoan12} Padoan, P., Haugb\o lle, T., Nordlund, \AA., 2012, \apj, 759, L27
\bibitem[Pellerin et al.(2008)]{pellerin08}]{} Pellerin, A., Meyer, M., Harris, J., \&
    Calzetti, D. 2008, ASPC, 388, 379
\bibitem[Radburn-Smith et al.(2012)]{radburn12}  Radburn-Smith, D.\ J., Ro\u{s}kar, R.,
    Debattista, V.\ P., et al.\ 2012, \apj, 753, 138
\bibitem[Ro\u{s}kar et al.(2008)]{roskar08} Ro\u{s}kar, R., Debattista, V. P., Quinn, T. R., Stinson, G. S., \& Wadsley, J. 2008,  \apj, 684, L79a
\bibitem[Rubio et al.(2015)]{rubio15} Rubio, M., Elmegreen, B.G., Hunter, D.A., Brinks, E.,
    Cort\'es, J.R., \& Cigan, P. 2015, Nature, 525, 218
\bibitem[Saha et al.(2010)]{saha10} Saha, A., Olszewski, E.\ W., Brondel, B.\
%Olsen, K., Knezek, P., Harris, J., Smith, C., Subramaniam, A., Claver, J., Rest, A. Seitzer, P., Cook, K. H., Minniti, D., \& Suntzeff, N. B. 2010, \aj, 140, 1719
\bibitem[S\'anchez Almeida et al.(2013)]{jorge13} S\'anchez Almeida, J., Mu\~noz-Tu\~n\'on,
    C. Elmegreen, D., Elmegreen, B., \& Mendez-Abreu, J. 2013, ApJ, 767, 74
\bibitem[S\'anchez Almeida et al.(2014)]{jorge14} S\'anchez Almeida, J. Morales-Luis, A.B.,
    Mu\~noz-Tu\~n\'on, C., Elmegreen, D.M.,  Elmegreen B.G., \& Mendez-Abreu, J. 2014a, ApJ,
    783, 45
\bibitem[S\'anchez Almeida et al.(2015)]{jorge15} S\'anchez Almeida, J. Elmegreen, B.G.,
    Munoz-Tunon, C., Elmegreen D.M., Perez-Montero, E., Amorin, R., Filho, M.E., Ascasibar,
    Y., Papaderos, P., \& Vilchez, J.M. 2015, ApJL, 810L, 15
\bibitem[Salpeter(1955)]{salpeter55} Salpeter, E. E. 1955, \apj, 121, 161
\bibitem[Toomre(1964)]{toomre64} Toomre, A. 1964, \apj, 139, 1217
\bibitem[van den Bergh(1988)]{vdbergh88} van den Bergh, S. 1988, PASP, 100, 344
\bibitem[Walter et al.(2008)]{walter08} Walter, F., Brinks, E., de Blok, W. J. G., et al. 2008, \aj, 136, 2563
\bibitem[Wang et al.(2014)]{wang14} Wang, J., Fu, J., Aumer, M., Kauffmann,
    G., J\'ozsa, G.I.G., Serra, P., Huang, M.-l., Brinchmann, J., van der Hulst, T., Bigiel, F. 2014
    MNRAS, 441, 2159
\bibitem[Wyder et al.(2007)]{wyder07} Wyder, T.\ K., et al.\ 2007, \apjs, 173, 293
\bibitem[Zhang et al.(2012)]{zhang12} Zhang, H.-X., Hunter, D. A., Elmegreen, B. G., Gao, Y., \& Schruba, A. 2012, \aj, 143, 47
\end{thebibliography}

\end{document}